# Carrier-envelope phase control of ultrafast photocurrents in layered MoS$_2$

J. Schmuck[1,2], B. Sinz[1,2], N. Pettinger[1,2], S. Zherebtsov[1,2] and A.W. Holleitner[1,2,*]

[1] *Walter Schottky Institute and Physics Department, Technical University of Munich, Am Coulombwall 4a, 85748 Garching, Germany.*

[2] *Munich Center of Quantum Science and Technology (MCQST), Schellingstraße 4, 80799 Munich, Germany.*

* holleitner@wsi.tum.de

**Abstract:**

We demonstrate carrier-envelope-phase (CEP)-controlled photocurrents in mono-, bi-, and tri-layer MoS$_2$ driven by few-cycle laser pulses. The photocurrent in the two-terminal devices scales quadratically with the field amplitude, indicating perturbative carrier dynamics in the weak-field regime distinct from strong-field tunnelling. Our results extend light-field-sensitive current control from bulk dielectrics, semiconductors, and graphene to two-dimensional transition-metal dichalcogenides, highlighting their potential for electric-field sensitive optoelectronics.

## 1. Introduction

The control of electron dynamics on attosecond to femtosecond timescales using the electric field of light has opened new directions for ultrafast electronics, enabling carrier dynamics to be manipulated on the same timescale as the optical waveform itself [1–3]. In this light-field-driven regime, the instantaneous electric field - rather than the cycle-averaged intensity - dictates the transport, injection, and acceleration of charge carriers in electronic devices [1,4,5]. Such control offers the potential to push signal processing, logic operations, and information transfer into the petahertz domain [6–8], linking optical physics with high-speed electronics [9,10]. Over the past decade, light-field-driven currents have been observed in a variety of materials, from conductors such as graphene [1,11] to bulk semiconductors [4] and wide-bandgap dielectrics [5,12], revealing a rich interplay between material properties, excitation conditions, and field strength. In the strong-field regime, tunnelling and interband transitions dominate, while for weak electric fields, the charge carrier response can be described in a perturbative approach [13].

Among the most promising material platforms for advancing nanoscale optoelectronics are transition-metal dichalcogenides (TMDCs) [14–17]. These atomically thin semiconductors combine strong light–matter interaction with a layered structure that allows their symmetry properties to be tuned by controlling the number of layers. Monolayer MoS$_2$ is inherently inversion asymmetric, making it an ideal system to study light-field-driven phenomena in a non-centrosymmetric crystal lattice. Adding a second layer restores inversion symmetry, while further layers typically increase the dielectric screening and modify the electronic band structure [18]. Recent work has demonstrated that light-field and spin-orbit-driven currents in van der Waals materials provide powerful routes for engineering ultrafast photocurrent responses that go beyond conventional photovoltaic and photoconductive mechanisms [8,19].

A key control parameter in the few-cycle laser excitation is the carrier-envelope phase (CEP), which determines the precise temporal waveform of the electric field [20–22]. In inversion-asymmetric materials, CEP control enables sub-cycle steering of carriers through the combined influence of the crystal symmetry and the optical field asymmetry [23].

Shifting the CEP by half a cycle inverts the driving field, which in turn reverses the direction of carrier motion and changes the sign of the photocurrent. Extensive work has clarified many aspects of CEP effects, revealing phenomena such as sub-cycle electron emission in strong-field ionization [20,24], CEP-dependent high-harmonic generation [25], and directional asymmetries in photoelectron momentum distributions [26]. Yet their behavior in 2D semiconductors remains largely unexplored, especially in the perturbative regime [13]. In the latter regime, the photocurrent does not arise from nonlinear tunneling, as in the strong-field case, but instead, it follows from field-driven separation and acceleration of charge carriers.

In this letter, we report the generation of CEP-dependent photocurrents in monolayer, bilayer, and trilayer $MoS_2$ in the weak-field regime. The measurements reveal a quadratic scaling of the photocurrent with the applied field strength, confirming the perturbative nature of the carrier dynamics. By exploring devices with different layer numbers, we show how the 2D geometry and the presence or absence of inversion symmetry govern the strength and polarity of the CEP response. Our findings extend the field of light-field-driven transport to atomically thin semiconductors as TMDCs with tunable symmetry, providing new opportunities for the design of ultrafast, low-dimensional optoelectronic devices operating on sub-cycle timescales.

## 2. Experimental scheme

To study the ultrafast electronic response of layered $MoS_2$, we irradiate the sample with a train of few-cycle CEP-stabilized laser pulses (cf. Fig. 1a). The optical pulses have a duration of 5.9 fs at a center wavelength of 826 nm (SI Fig. S1), pulse energies up to 2.2 nJ, and a repetition frequency $f_{rep}$ = 80 MHz. The carrier-envelope offset frequency ($f_{CEO}$) is stabilized by an f-2f feedback loop. We pre-compress the pulses with a pair of chirped mirrors to compensate for dispersion in the optical path and then pass them through a variable neutral density wheel to attenuate the laser power. For lock-in amplification and detection of the photocurrents, the laser beam is chopped at a reference frequency $f_{chopper}$ = 327 Hz. Finally, the beam passes a pair of $CaF_2$ wedges before it is focused on the sample with the help of an off-axis parabolic mirror.

For the samples, we mechanically exfoliate mono- (1L-), bi- (2L-), and tri-layer (3L-) $MoS_2$ from a $MoS_2$ bulk crystal and transfer them onto strongly $p$-doped Si substrates with a 285 nm-thick $SiO_2$-layer on top. The samples are characterized by Raman and photoluminescence spectroscopy to confirm the layer number and quality of the $MoS_2$ layers (cf. SI Fig. S2). By optical lithography and e-beam evaporation, two-terminal devices with parallel contacts are fabricated. The metal contacts are made from Ti and Au (5 nm / 100 nm), and the devices show a linear current-voltage characteristic at small voltage bias (SI Fig. S3). For the photocurrent measurements, we position the samples in a vacuum chamber at a pressure of $2 \times 10^{-5}$ mbar. The beam diameter is approximately 5 µm at the sample plane (see SI Fig. S4). The linearly polarized laser aligns perpendicular to the electrodes, i.e. along the current path between the contacts. The setup allows scanning the sample in all three spatial dimensions. By translating one of the $CaF_2$ wedges perpendicular to the beam path, we introduce an additional optical path $\Delta D_{wedge}$, which controls the carrier-envelope phase (CEP) of the laser pulses. This adjustment modulates the photocurrent in $MoS_2$ according to the resulting phase shift $\Delta\phi_{CEP}$ (cf. Fig. 1a).

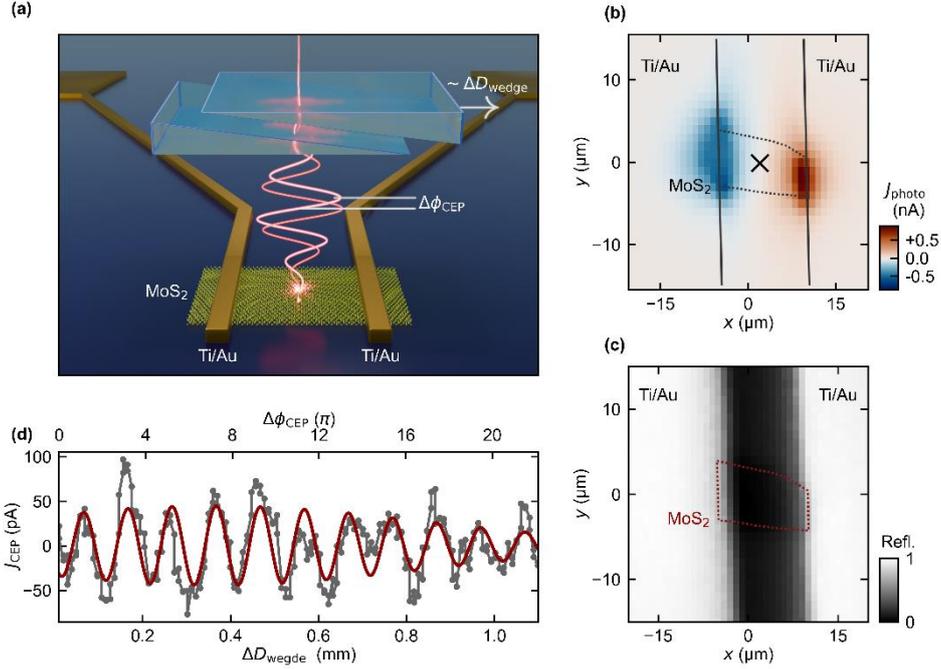

**Fig. 1.** Carrier-envelope phase (CEP) - dependent photocurrents in monolayer MoS$_2$. (a) Schematic illustration of the studied device. The CEP of the laser pulse train is changed by a distance $\Delta D_{\text{wedge}}$ by translating one of the two utilized CaF$_2$ wedges. The photocurrent in MoS$_2$ is modulated by the corresponding phase change $\Delta \phi_{\text{CEP}}$. (b) Spatial map of the time-integrated photocurrent $J_{\text{photo}}$ of a two-terminal 1L-MoS$_2$ device at $V_{\text{SD}} = 0$V. The black cross indicates a central position on the monolayer. (c) Spatial reflectance map of the device indicating the position of the contacts. Dashed red line highlights the extensions of the 1L-MoS$_2$. (d) A plot of the CEP-dependent photocurrent contribution $J_{\text{CEP}}$ vs. the wedge insertion $\Delta D_{\text{wedge}}$ as measured at the cross in (b).

With the given geometry of the wedges, we calculate a phase shift of $2\pi$ by an additional insertion of

$$D_{\text{CEP}} = \left| \frac{\partial n_{\text{CaF}_2}}{\partial \lambda} \right|_{\lambda_0=826\text{nm}}^{-1} \approx 102.9 \, \mu m.$$

All the experiments are performed at zero bias voltage ($V_{\text{SD}} = 0$V) at room temperature.

### 3. Results

We observe extrema of the time-integrated photocurrent $J_{\text{photo}}$ close to the contacts (cf. Fig. 1b), the position of which is determined by concurrently measured reflectance maps (Fig. 1c). The two distinct extrema have opposite signs and they can be understood to originate from a combination of built-in electric fields at the MoS$_2$ / metal interfaces and hot charge carriers as driven by a photo-thermoelectric effect [15]. The black cross in Figure 1b marks the position where we carry out the CEP modulation measurements. Figure 1d presents the corresponding measurement of the CEP-sensitive photocurrent contribution $J_{\text{CEP}}$. By translating the glass wedge and thus varying the phase over several multiple of $2\pi$, the signal oscillates with the characteristic periodicity length $D_{\text{CEP}}$. The modulation can be understood as follows: For a cosine shaped pulse with $\phi_{\text{CEP}} = 0$, the asymmetry of the laser pulse is strongest, giving rise to the maximum $J_{\text{CEP}}$. A phase of $\pi/2$ eliminates the asymmetry, causing a zero crossing of $J_{\text{CEP}}$ with wedge insertion, and at $\pi$ the effect is reversed, producing a minimum [6]. The modulations are usually on top of a large photocurrent background originating from intensity-dependent mechanisms (cf. Fig. 2).

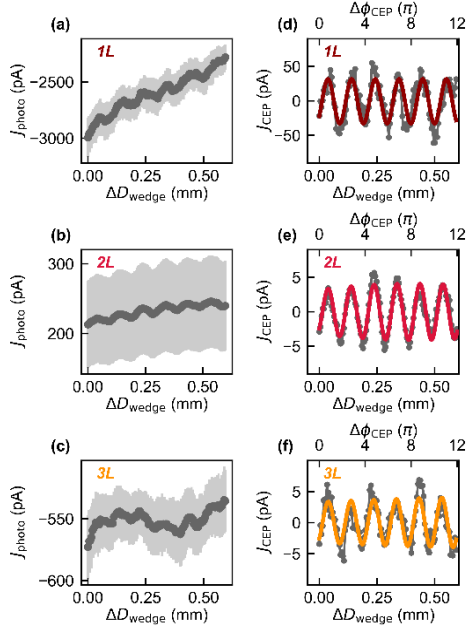

**Fig. 2.** CEP-sensitive photocurrents in 1L-, 2L-, and 3L-MoS$_2$ devices. (a) Total photocurrent signal $J_\text{photo}$ vs. $\Delta D_\text{wedge}$ for a 1L-MoS$_2$ with reproducible oscillations. Light gray area highlights the standard deviation of several measurements. (b) and (c): Similar measurements for devices with a 2L-MoS$_2$ (b) and a 3L-MoS$_2$ (c). (d)-(f): The CEP-dependent photocurrent contributions for the 1L-MoS$_2$ (d), 2L-MoS$_2$ (e), and 3L-MoS$_2$ (f) obtained by subtracting the background contributions from the $J_\text{photo}$ in (a), (b), and (c).

Figures 2a-c depict the CEP-dependent scans of $J_\text{photo}$ vs. $\Delta D_\text{wedge}$ as measured with two-terminal devices based on 1L- (a), 2L- (b), and 3L- (c) MoS$_2$ flakes. The measurements are performed at the center of each device (cf. Fig. 1b and SI Fig. S5) [2]. For each device, multiple scans are recorded and the corresponding average value of $J_\text{photo}$ is presented with a light gray error bar indicating the standard deviation across the measurements. By subtracting a higher-order polynomial from $J_\text{photo}$, we extract the CEP-dependent contribution $J_\text{CEP}$ (cf. gray data in Figs. 2d-f). Then, we fit the data with a function assuming a Gaussian envelope due to spectral dispersion in the CaF$_2$ wedges [1]:

$$J_\text{CEP} = J_{\text{CEP},0} \cdot \exp\left[-\frac{(\Delta D_\text{wedge} - D_0)^2}{2\sigma_\text{CEP}^2}\right] \cdot \cos(2\pi \cdot \Delta D_\text{wedge}/D_\text{CEP} - \phi_j)$$

$$= J_{\text{CEP},0} \cdot \exp\left[-\frac{(\Delta D_\text{wedge} - D_0)^2}{2\sigma_\text{CEP}^2}\right] \cdot \cos(\Delta\phi_\text{CEP} - \phi_j). \quad (1)$$

In these fits (cf. lines in Figs. 2d-f), $J_{\text{CEP},0}$ represents the maximum amplitude at the wedge insertion offset $D_0$, and $\sigma_\text{CEP}$ denotes the standard deviation of the CEP-dependent signal envelope. Furthermore, $\phi_j$ corresponds to the absolute phase offset of the oscillating signal. Generally, equation (1) describes an oscillating signal with a maximum at $D_0$, when the pulse duration is shortest and the electric field is strongest. The periodicity of the introduced modulation is the same for all shown measurements. Moreover, the oscillation amplitude $J_{\text{CEP},0}$ is largest for the 1L-MoS$_2$ and it is significantly smaller for the 2L- and 3L-MoS$_2$ (cf. Figs. 2d- f).

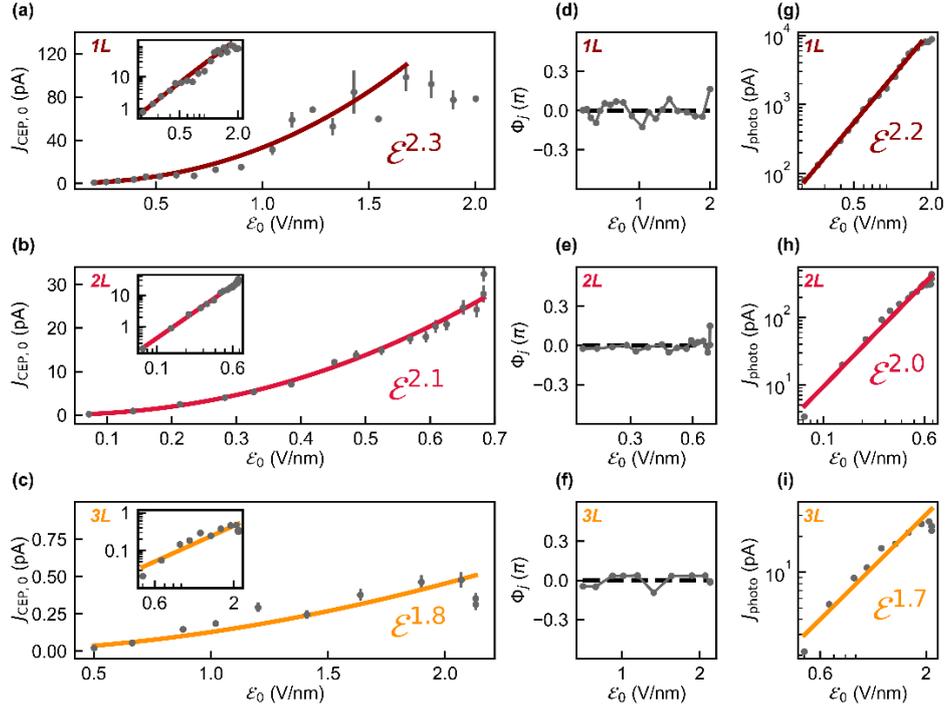

**Fig. 3**. Laser field-dependence of the CEP-photocurrents for the 1L-, 2L-, and 3L-MoS$_2$. (a-c): The fitted amplitude $J_{CEP,0}$ as a function of the peak electric field strength $\mathcal{E}_0$ for (a) 1L-, (b) 2L-, and (c) 3L-MoS$_2$. Lines are fits to the data. Respective insets show the data and the fits in a double-logarithmic representation. (d-f): All data points in (a) – (c) show a stable phase $\phi_j$ as a function of $\mathcal{E}_0$. (g-i): Double-logarithmic plot of the time-integrated photocurrent $J_{photo}$ vs. $\mathcal{E}_0$ exhibiting a similar scaling as $J_{CEP,0}$ vs. $\mathcal{E}_0$. Lines are fits to the data.

Figures 3a-c show the scaling of the $J_{CEP,0}$ with the peak electric field $\mathcal{E}_0$ of the driving laser pulse for the 1L-, 2L-, and 3L-MoS$_2$. For all three samples, the amplitude $J_{CEP,0}$ increase monotonically with the field strength up to a threshold value of 1.7 V/nm for 1L-, 0.65 V/nm for 2L-, and 2.1 V/nm for 3L-MoS$_2$. At values above these ranges, we attribute the signal saturation to a visible damage to the samples, reflecting their damage threshold under illumination.

Fitting the data in the below-threshold regions reveals a near quadratic dependence on the peak laser field for all three samples. Further analysis of the $J_{CEP}$ shows a near constant photocurrent phase $\phi_j$ dependence on the driving laser field strength (see Figs. 3d-f). Therefore, within the investigated range of peak electric fields, we do not observe any current reversals [1]. Moreover, the scaling of the total photocurrents $J_{photo}$ with field strength shows also an approximately quadratic dependence (Figs. 3g-i).

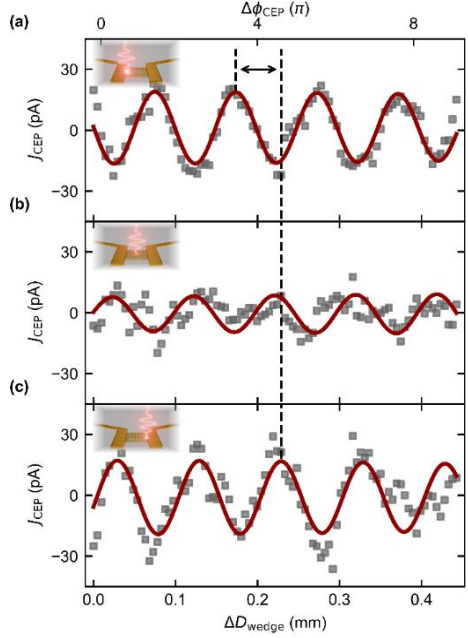

**Fig. 4.** Spatial dependences of the CEP dependent photocurrents contributions of the 1L-MoS$_2$ device. (a) $J_{CEP}$ vs. $\Delta D_{wedge}$ at the left contact. (b) Similar data at the center of the monolayer. (c) $J_{CEP}$ vs. $\Delta D_{wedge}$ at the right contact. Vertical line demonstrates a phase change of π of $J_{CEP}$ when measured at the left and right contact.

Finally, we explore the influence of the MoS$_2$/contact interfaces on generation of the CEP-dependent photocurrents. To this end, we illuminate the 1L-MoS$_2$ sample of Fig. 1 at various positions along the *x*-axis, i.e. from one of the two contacts across the center to the other contact. Figure 4 presents three exemplary traces measured with an illumination of (a) the left electrode, (b) the center of the flake, and (c) the right electrode. From comparison of the signals, two main trends can be identified. First, illumination of the MoS$_2$-contact interface results in higher current amplitudes as compared to excitation in the center of the flake. Second, the oscillations in (a) and (c) are almost out of phase by π, and the phase of the oscillating signal at the center position (b) lies in-between (cf. vertical dashed line and SI Fig. S7).

### 4. Discussion and conclusion

Our experimental results demonstrate that, similar to other 2-D and bulk materials, excitation with waveform-controlled few-cycle laser fields can induce CEP-dependent photocurrents in layered MoS$_2$ devices. The CEP modulation of the observed current, periodic with wedge insertion (cf. Fig. 1d), follows the expected sinusoidal dependence. The CEP-dependent contributions are strongest in the monolayer case (1L-MoS$_2$), which we attribute to the increased light-matter interaction because of a direct band gap transition. For 2L-MoS$_2$ and 3L-MoS$_2$, the optical transitions become increasingly indirect [18]. Moreover, we observe a near quadratic dependence of the CEP-controlled signal on the incident laser field (cf. Fig. 3), which indicates that single-photon excitation dominates the current generation process. This finding is in contrast to previous work on CEP-dependent photocurrents where a 3$^{rd}$ and higher order power law scaling was observed; attributed to a coherent superposition of out-of-phase multiphoton electronic excitation [27,28] and/or strong field inter- and intra-band dynamics [1,4,12]. Instead, our results can be understood in terms of the weak-field or perturbative regime [13]. Consistently, we observe that the quadratic dependence and the phase of the CEP-dependent oscillations remains the same as a function of the driving field strength (cf. Figures 3a–f). Numerical estimations (cf. SI) suggest that similar to ref. [29], a single-photon excitation in a direct band-gap transition in combination with the waveform-controlled steering of the photoexcited electrons can result in a non-negligible CEP-dependence of the residual current (cf. SI Fig. S8). As the laser spectrum extends to 600 nm (~2.06 eV), we expect that the A and B excitonic states of $MoS_2$ (at ~1.84 eV and ~2 eV in the investigated samples) can be directly populated with the laser pulse (cf. SI Figs. S1 and S2). Moreover, it has been shown for monolayer $MoS_2$, that the exciton formation can be considered to occur on the few-femtosecond timescale scale [30]. Hence, we assume that the resonant excitation and subsequent dissociation of the excitons can happen within the femtosecond duration of the utilized laser pulse. Once an electron is promoted from the valence band to the conduction band, it acquires the final momentum proportional to the vector potential of the driving field at the moment of the transition. Within this steering approach, we estimate the CEP-modulation of the electron momentum distribution in the conduction band to be on the order of $10^{-4}$ (cf. SI Fig. S8). From the experimental data in Fig. 3, we estimate the ratio of $J_{CEP}/J_{photo}$ to be in the order of $10^{-2}$. We note, however, that in the calculus, we do not consider any CEP-dependent contribution from residual charge carriers in the conduction band, as this would require further consideration of the nonlinear conduction-band dynamics [10], which is beyond the scope of the current work. Still, the calculus together with the similarity of the power scaling of $J_{CEP}$ and $J_{photo}$ (cf. Fig. 3), prompts us to tentatively ascribe the observed CEP-dependent signal to a weak CEP sensitivity of the conventional mechanisms, such as photovoltaic or photo-thermoelectric currents [15]. We note that this interpretation is also consistent with the findings on the spatial dependence of the photocurrent (cf. Figs 1b and 4). There, both the CEP-dependent and the total photocurrent signals exhibit a stronger response at the metal–$MoS_2$ interface than compared to an excitation in the center of the $MoS_2$ device. Remarkably, this spatial dependence is different from previous work on graphene in the strong-field regime, where the spatial dependence of the CEP-dependent signal amplitude seems to have an opposite trend [2]. Interestingly, similar to the previous work [2], we observe a variation of the phase $\phi_j$ as the optical focus is moved from one electrode to the other. In our interpretation, this behavior can be attributed to Schottky barriers formed at the metal–$MoS_2$ interfaces in combination with the optical laser field. In a simple picture, the laser field transiently distorts the barriers and either promotes or hinders the transfer of photoexcited carriers into the metal contacts, depending on the instantaneous field direction [3,31]. Moreover, the specific geometry and morphology of the metal–$MoS_2$ interfaces might further impact the photocurrent generation mechanism on the femtosecond timescale [15,32].

In conclusion, our results demonstrate that CEP-controlled currents can be generated in layered $MoS_2$-based two-terminal devices. We find a quadratic scaling of the corresponding amplitude similar to the scaling behavior of the background photocurrent stemming from either photo-thermoelectric or photovoltaic background processes. We interpret our findings in a way that a CEP-dependent steering of charge carriers after their excitation leads to a small, but measurable quadratically scaling CEP-dependent current. The polarity of the current is reversed at the contact interfaces, where the time- and spatial-symmetry is broken. Our study reveals CEP-sensitive currents in the weak-field regime in the two-dimensional materials, excluding any strong-field physics. On a broader perspective, our results open pathways for using the contact's geometry to generate tailored current responses. Particularly, integrating plasmonic or dielectric nanostructures and tailoring of the properties of the 2D materials could provide a route to stronger CEP currents at even lower incident powers. Looking ahead, our findings suggest that two-dimensional van der Waals semiconductors offer a versatile base for developing field-sensitive optoelectronic devices that operate at optical-cycle timescales. The demonstrated robustness of CEP control also highlights technological relevance for ultrafast photodetection, waveform sampling, and on-chip signal processing. More broadly, the ability to steer and excite charge carriers while controlling the residual current by external measures as the CEP in the weak-field regime contributes to the field of petahertz electronics, where charge transport is dictated not by scattering times but directly by the optical field.

**Funding.** We gratefully acknowledge financial support by the Munich Quantum Valley (K1) and the DFG via project DFG HO3324 / 13 and the excellence clusters Munich Center for Quantum Science and Technology (MCQST) and e-conversion.

**Acknowledgements**. We thank C. Kastl for helpful discussions and U. Kleineberg for the instrumental support.

**Conflict of Interest**. The authors have no conflicts to disclose.

**Data availability.** The data that support the findings of this study are available from the corresponding authors upon reasonable request.

**References**

1. T. Higuchi, C. Heide, K. Ullmann, H. B. Weber, and P. Hommelhoff, "Light-field-driven currents in graphene," Nature **550**, 224–228 (2017).
2. T. Boolakee, C. Heide, A. Garzón-Ramírez, H. B. Weber, I. Franco, and P. Hommelhoff, "Light-field control of real and virtual charge carriers," Nature **605**, 251–255 (2022).
3. T. Rybka, M. Ludwig, M. F. Schmalz, V. Knittel, D. Brida, and A. Leitenstorfer, "Sub-cycle optical phase control of nanotunnelling in the single-electron regime," Nature Photon **10**, 667–670 (2016).
4. F. Langer, Y.-P. Liu, Z. Ren, V. Flodgren, C. Guo, J. Vogelsang, S. Mikaelsson, I. Sytcevich, J. Ahrens, A. L'Huillier, C. L. Arnold, and A. Mikkelsen, "Few-cycle lightwave-driven currents in a semiconductor at high repetition rate," Optica, OPTICA **7**, 276–279 (2020).
5. A. Schiffrin, T. Paasch-Colberg, N. Karpowicz, V. Apalkov, D. Gerster, S. Mühlbrandt, M. Korbman, J. Reichert, M. Schultze, S. Holzner, J. V. Barth, R. Kienberger, R. Ernstorfer, V. S. Yakovlev, M. I. Stockman, and F. Krausz, "Optical-field-induced current in dielectrics," Nature **493**, 70–74 (2013).
6. P. D. Keathley, W. P. Putnam, P. Vasireddy, R. G. Hobbs, Y. Yang, K. K. Berggren, and F. X. Kärtner, "Vanishing carrier-envelope-phase-sensitive response in optical-field photoemission from plasmonic nanoantennas," Nat. Phys. **15**, 1128–1133 (2019).
7. M. Borsch, M. Meierhofer, R. Huber, and M. Kira, "Lightwave electronics in condensed matter," Nat Rev Mater **8**, 668–687 (2023).
8. C. Heide, P. D. Keathley, and M. F. Kling, "Petahertz electronics," Nat Rev Phys **6**, 648–662 (2024).
9. M. Ossiander, K. Golyari, K. Scharl, L. Lehnert, F. Siegrist, J. P. Bürger, D. Zimin, J. A. Gessner, M. Weidman, I. Floss, V. Smejkal, S. Donsa, C. Lemell, F. Libisch, N. Karpowicz, J. Burgdörfer, F. Krausz, and M. Schultze, "The speed limit of optoelectronics," Nat Commun **13**, 1620 (2022).
10. C. Karnetzky, P. Zimmermann, C. Trummer, C. Duque Sierra, M. Wörle, R. Kienberger, and A. Holleitner, "Towards femtosecond on-chip electronics based on plasmonic hot electron nano-emitters," Nat Commun **9**, 2471 (2018).
11. B. Fehér, V. Hanus, W. Li, Z. Pápa, J. Budai, P. Paul, A. Szeghalmi, Z. Wang, M. F. Kling, and P. Dombi, "Light field–controlled PHz currents in intrinsic metals," Science Advances **11**, eadv5406 (2025).
12. V. Hanus, V. Csajbók, Z. Pápa, J. Budai, Z. Márton, G. Z. Kiss, P. Sándor, P. Paul, A. Szeghalmi, Z. Wang, B. Bergues, M. F. Kling, G. Molnár, J. Volk, and P. Dombi, "Light-field-driven current control in solids with pJ-level laser pulses at 80 MHz repetition rate," Optica, OPTICA **8**, 570–576 (2021).
13. C. Heide, T. Boolakee, T. Higuchi, and P. Hommelhoff, "Adiabaticity parameters for the categorization of light-matter interaction: From weak to strong driving," Phys. Rev. A **104**, 023103 (2021).
14. H. Wang, C. Zhang, W. Chan, S. Tiwari, and F. Rana, "Ultrafast response of monolayer molybdenum disulfide photodetectors," Nat Commun **6**, 8831 (2015).
15. E. Parzinger, M. Hetzl, U. Wurstbauer, and A. W. Holleitner, "Contact morphology and revisited photocurrent dynamics in monolayer MoS2," npj 2D Mater Appl **1**, 40 (2017).
16. C. Heide, T. Boolakee, T. Higuchi, and P. Hommelhoff, "Sub-cycle temporal evolution of light-induced electron dynamics in hexagonal 2D materials," J. Phys. Photonics **2**, 024004 (2020).
17. S. A. Oliaei Motlagh, V. Apalkov, and M. I. Stockman, "Transition metal dichalcogenide monolayers in an ultrashort optical pulse: Femtosecond currents and anisotropic electron dynamics," Phys. Rev. B **103**, 155416 (2021).
18. K. F. Mak, C. Lee, J. Hone, J. Shan, and T. F. Heinz, "Atomically Thin MoS2: A New Direct-Gap Semiconductor," Phys. Rev. Lett. **105**, 136805 (2010).
19. J. Kiemle, P. Zimmermann, A. W. Holleitner, and C. Kastl, "Light-field and spin-orbit-driven currents in van der Waals materials," Nanophotonics **9**, 2693–2708 (2020).
20. G. G. Paulus, F. Grasbon, H. Walther, P. Villoresi, M. Nisoli, S. Stagira, E. Priori, and S. De Silvestri, "Absolute-phase phenomena in photoionization with few-cycle laser pulses," Nature **414**, 182–184 (2001).
21. D. J. Jones, S. A. Diddams, J. K. Ranka, A. Stentz, R. S. Windeler, J. L. Hall, and S. T. Cundiff, "Carrier-Envelope Phase Control of Femtosecond Mode-Locked Lasers and Direct Optical Frequency Synthesis," Science **288**, 635–639 (2000).

22. E. Goulielmakis, M. Uiberacker, R. Kienberger, A. Baltuska, V. Yakovlev, A. Scrinzi, Th. Westerwalbesloh, U. Kleineberg, U. Heinzmann, M. Drescher, and F. Krausz, "Direct Measurement of Light Waves," Science **305**, 1267–1269 (2004).
23. C. Heide, T. Boolakee, T. Eckstein, and P. Hommelhoff, "Optical current generation in graphene: CEP control vs. ω + 2ω control," Nanophotonics **10**, 3701–3707 (2021).
24. G. Hergert, R. Lampe, A. Wöste, and C. Lienau, "Ultra-Nonlinear Subcycle Photoemission of Few-Electron States from Sharp Gold Nanotapers," Nano Lett. **24**, 11067–11074 (2024).
25. A. Baltuška, T. Udem, M. Uiberacker, M. Hentschel, E. Goulielmakis, C. Gohle, R. Holzwarth, V. S. Yakovlev, A. Scrinzi, T. W. Hänsch, and F. Krausz, "Attosecond control of electronic processes by intense light fields," Nature **421**, 611–615 (2003).
26. M. F. Kling, Ch. Siedschlag, A. J. Verhoef, J. I. Khan, M. Schultze, Th. Uphues, Y. Ni, M. Uiberacker, M. Drescher, F. Krausz, and M. J. J. Vrakking, "Control of Electron Localization in Molecular Dissociation," Science **312**, 246–248 (2006).
27. A. Haché, Y. Kostoulas, R. Atanasov, J. L. P. Hughes, J. E. Sipe, and H. M. van Driel, "Observation of Coherently Controlled Photocurrent in Unbiased, Bulk GaAs," Phys. Rev. Lett. **78**, 306–309 (1997).
28. R. Atanasov, A. Haché, J. L. P. Hughes, H. M. van Driel, and J. E. Sipe, "Coherent Control of Photocurrent Generation in Bulk Semiconductors," Phys. Rev. Lett. **76**, 1703–1706 (1996).
29. N. Altwaijry, M. Qasim, M. Mamaikin, J. Schötz, K. Golyari, M. Heynck, E. Ridente, V. S. Yakovlev, N. Karpowicz, and M. F. Kling, "Broadband Photoconductive Sampling in Gallium Phosphide," Advanced Optical Materials **11**, 2202994 (2023).
30. C. Trovatello, F. Katsch, N. J. Borys, M. Selig, K. Yao, R. Borrego-Varillas, F. Scotognella, I. Kriegel, A. Yan, A. Zettl, P. J. Schuck, A. Knorr, G. Cerullo, and S. D. Conte, "The ultrafast onset of exciton formation in 2D semiconductors," Nat Commun **11**, 5277 (2020).
31. P. Zimmermann, A. Hötger, N. Fernandez, A. Nolinder, K. Müller, J. J. Finley, and A. W. Holleitner, "Toward Plasmonic Tunnel Gaps for Nanoscale Photoemission Currents by On-Chip Laser Ablation," Nano Lett. **19**, 1172–1178 (2019).
32. T. Higuchi, L. Maisenbacher, A. Liehl, P. Dombi, and P. Hommelhoff, "A nanoscale vacuum-tube diode triggered by few-cycle laser pulses," Appl. Phys. Lett. **106**, 051109 (2015).

**SUPPLEMENTARY INFORMATION**

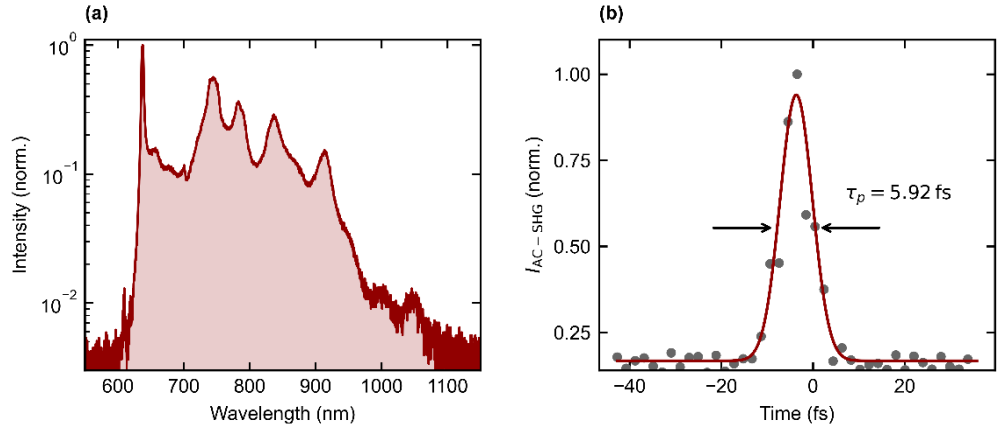

**Fig. S1.** Spectral density of the utilized optical pulses and the corresponding SHG-autocorrelation trace with a retrieved pulse length of 5.92 fs.

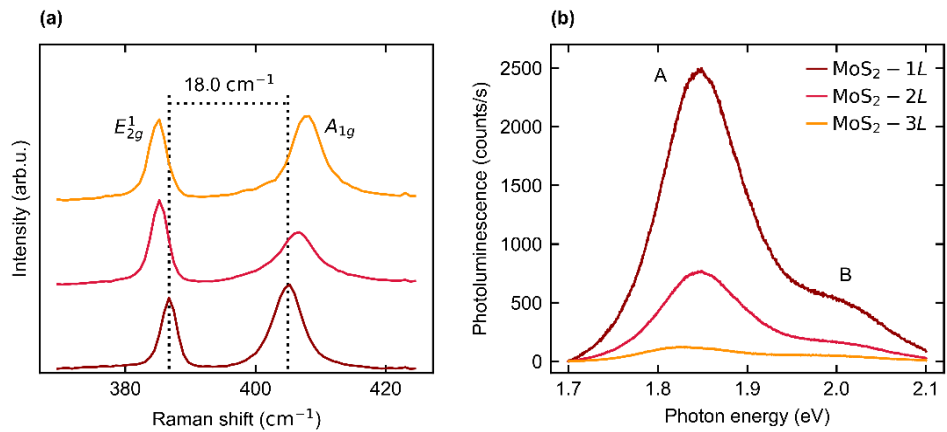

**Fig. S2.** (a) Spectrally resolved Raman- and (b) photoluminescence-spectra of the investigated $MoS_2$ flakes (measured at the center of the corresponding two-terminal devices). All measurements at room temperature.

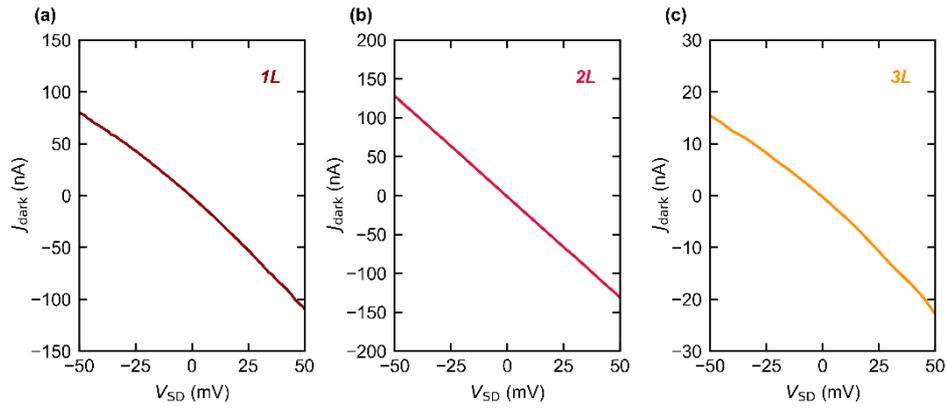

**Fig. S3.** *IV*-curves for the three samples. All measurements at room temperature.

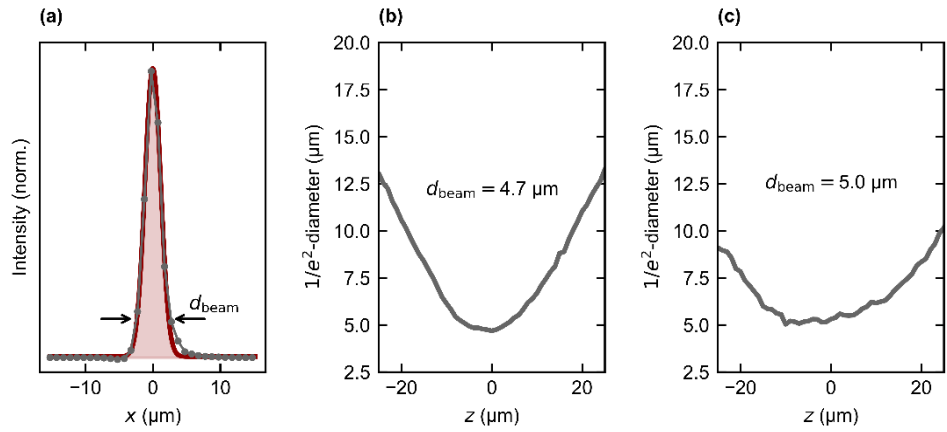

**Fig. S4.** Measurement of the laser focus. (a) Retrieved laser intensity profile in the horizontal direction at the focus position $z = 0$. (b) and (c): $1/e^2$-intensity diameter in the vertical and horizontal direction for positions along the $z$-axis (larger z values correspond to greater distances from the OAP mirror ).

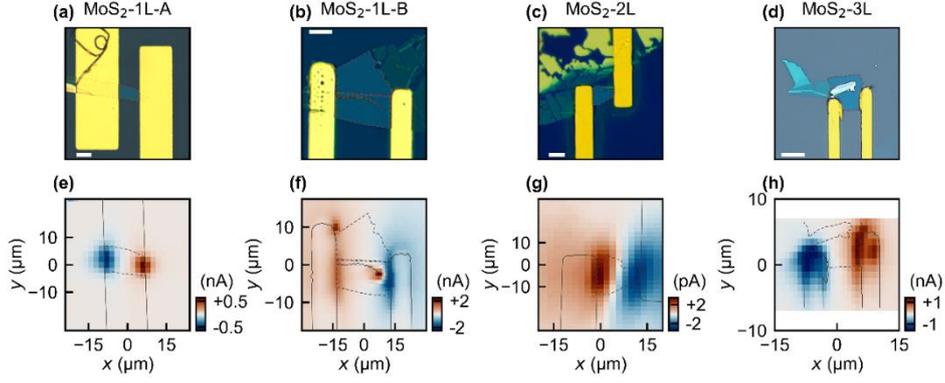

**Fig. S5.** Optical images of the devices (monolayer: a/b, bilayer: c, trilayer: d). The white scale bar is 10 µm. (e)-(h): Photocurrent maps of the respective samples. The colorbars indicate the magnitude of $J_{\text{photo}}$. All measurements at room temperature.

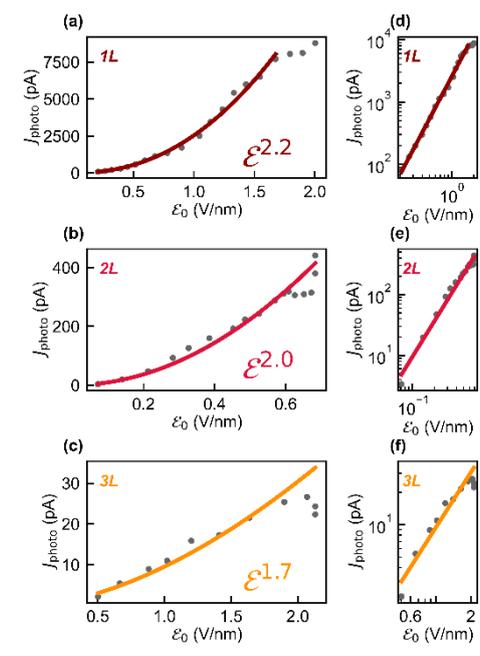

**Fig. S6.** Scaling of total photocurrents $J_{\text{photo}}$ for the (a) monolayer, (b) bilayer and (c) trilayer sample. Data are the same as the one in Fig. 3 of the main manuscript. (d)-(f): Double-logarithmic representations of the data with fits for power exponents. are extracted from the fit in the double-logarithmic representations (d-f), respectively.

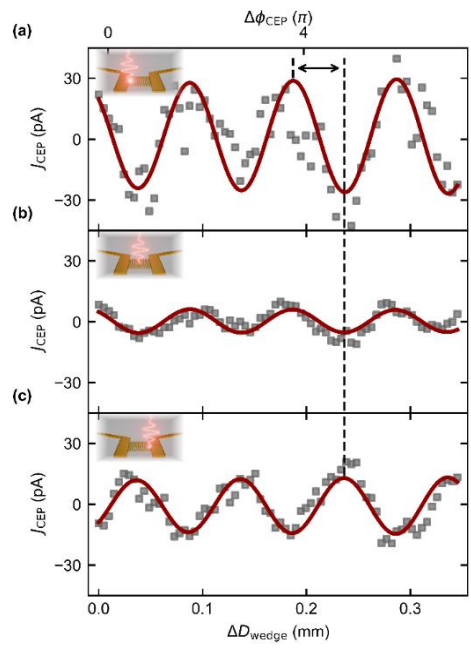

**Fig. S7.** Phase shift of CEP-dependent current on the two MoS$_2$-electrode interfaces in the monolayer reference sample (sample MoS$_2$-1L-B as in Fig. S5).

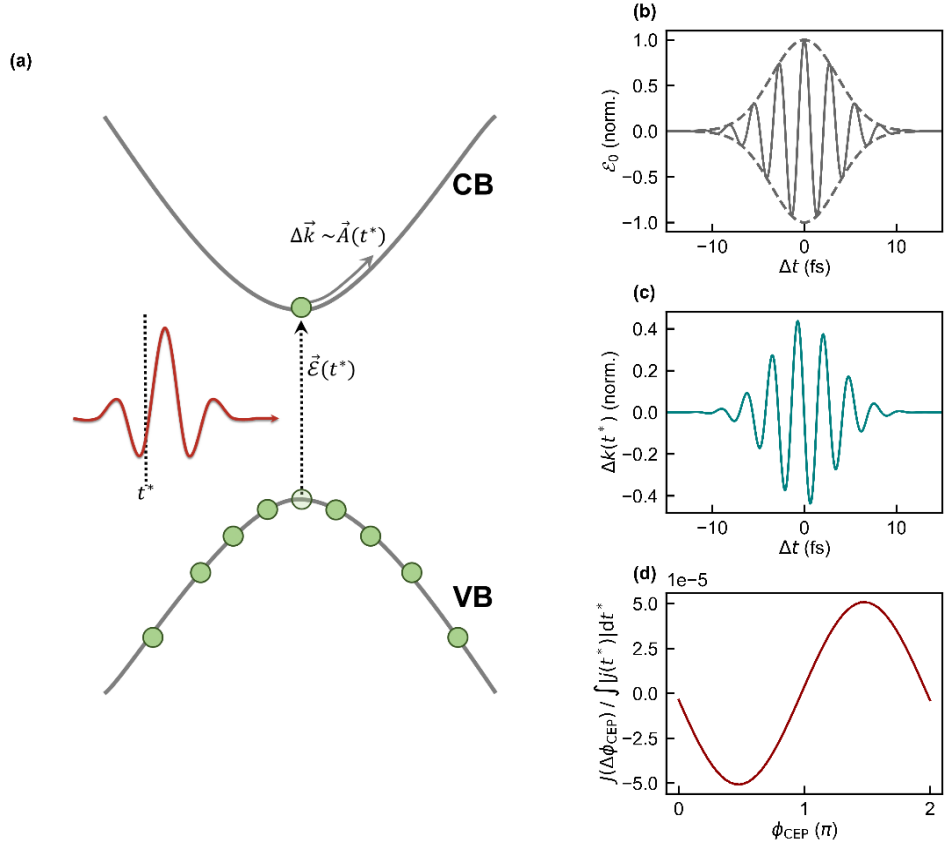

**Fig. S8.** (a) Illustration of the processes during and after light-matter interaction. (b) A few-cycle laser pulse induces single-photon transition from the valence to the conduction band proportional to the square of the envelope (dashed line). (c) Acquired momentum offset $\Delta k(t^*)$ from an electron excited at a time $t^*$. (d) CEP-dependent current contribution based on the independent-particle picture simulation.

The aim of the simulation in Fig. S8 is to model the generation of photocurrent as a function of the CEP of a few-cycle laser pulse, using a semiclassical approach. For the MoS$_2$ monolayer, the conduction-band effective mass is approximated as $m^* = 0.54\, m_\text{e}$. Optical excitation is described through an effective single-photon absorption coefficient $\alpha$, enabling the excitation probability to be expressed as a function of the electric field strength (Fig. S8a). Further, the laser pulse is modeled with a central wavelength $\lambda = 826$ nm and $\tau_p = 5.9$ fs. The temporal profile of the field is given by $\mathcal{E}(t) = \mathcal{E}_0 \cos(2\pi t/T + \Delta\phi_\text{CEP}) \exp(-t^2/(2\tau_p^2))$, where a$_0$ is the normalized peak amplitude and $\Delta\phi_\text{CEP}$ the carrier-envelope phase (see Fig. S8b). The semiclassical Bloch acceleration framework is used to follow the electron dynamics. The electron momentum evolves as

$$k(t^*) = k_0 - \left(\frac{e}{\hbar}\right) \int_{t^*}^{\infty} \mathcal{E}(t') \, dt',$$

which in turn defines the conduction-band group velocity (see Fig. S8c),

$$v(t^*) = \hbar \frac{k(t^*)}{m^*}.$$

The excitation at each instant is modeled as $W(t^*) = \alpha \, \mathcal{E}(t^*)^2$, which reflects single-photon contributions. The contribution to residual current from electrons ionized at the time $t^*$ is given by

$$J(t^*) = W(t^*) \, v(t^*),$$

and the net CEP-dependent current is obtained by integrating over the pulse duration,

$$J(\Delta\phi_{\text{CEP}}) = \int_{-\infty}^{\infty} J(t^*) \, dt^*.$$

By scanning the CEP over the range from 0 to $2\pi$, the resulting current exhibits a periodic dependence on the carrier-envelope phase (see Fig. S8d). The amplitude of this dependence is strongly influenced by the pulse asymmetry. We estimate the current stemming from this effect to be $\frac{J(\Delta\phi_{CEP})}{\int |j(t^*)| \, dt^*} \sim 10^{-4}$ (cf. Fig. S8d). Moreover, the experimental data suggests a quadratic scaling of the CEP-dependent contribution with the electric field. The deviations from the model can be explained by processes not considered: We expect that processes as broadband absorption, recombination, multi-particle Coulomb interactions and electron-electron scattering might change the scaling with the field. The outcome of the simulation is therefore a CEP-resolved photocurrent, consistent with experimental observations of strong-field induced currents in MoS$_2$, where the carrier-envelope phase acts as a control knob for ultrafast current generation.